\documentclass[letterpaper, 10 pt, journal]{IEEEtran} 

\IEEEoverridecommandlockouts                              
\usepackage{amsmath,mathrsfs}
\usepackage{amstext}
\usepackage{color,colortbl}
\usepackage[table]{xcolor}
\usepackage{amsfonts}
\usepackage{amssymb,amsbsy}
\usepackage{graphicx}
\usepackage{epstopdf}
\usepackage{url}
\usepackage[us]{datetime}
\usepackage{cite}
\usepackage[thmmarks, amsmath]{ntheorem}

\definecolor{Gray}{gray}{0.9}
\definecolor{LightCyan}{rgb}{0.88,1,1}
\definecolor{Azure}{rgb}{0.96,1,1}
\definecolor{LightMagenta}{rgb}{1,0.88,1}
\definecolor{LightOrange}{rgb}{1,0.99,0.88}
\definecolor{LightGreen}{rgb}{0.9,1,0.85}
\definecolor{DarkOrange}{rgb}{0.98,0.91,0.71}
\definecolor{DarkGreen}{rgb}{0.67,0.88,0.69}

\allowdisplaybreaks[1]
\graphicspath{{../}, {../Bio/}, {../Fig/}, {../Fig02/}, {../Fig03/}, {../Fig04/}, {../Fig05/}, {../Fig06/}, {../Fig07/},
{../Fig08/}, {../Fig09/}, {../Fig10/}, {../Fig11/}, {../Fig12/}, {../Fig13/}}

\theorembodyfont{\normalfont}
\theoremseparator{:}
\theoremstyle{nonumberplain}
\theoremsymbol{$\square$}

\newcommand{\bfx}[0]{\boldsymbol{x}}

\newcommand{\bfu}[0]{\boldsymbol{u}}
\newcommand{\bfA}[0]{\mathbf{A}}
\newcommand{\bfB}[0]{\mathbf{B}}
\newcommand{\bfI}[0]{\mathbf{I}}

\newcommand{\sat}[0]{\textrm{sat}}
\newcommand{\sign}[1]{\text{sign}\left(#1\right) }

\title{
A Planning-free Longitudinal Controller Design for Vehicles in Dynamic Traffic Environments
}
\IEEEpubid{{\color{red} This paper is currently under review. This preprint is dedicated for REVIEW ONLY}.}

\author{Wubing~B.~Qin
\thanks{Manuscript revised \currenttime, \today.%
}
\thanks{Wubing B.~Qin is with the Department of Mechanical Engineering, University of Michigan, Ann Arbor, MI 48109, USA. (Email: wubing@umich.edu,).}%
}
\begin{document}
\maketitle
\begin{abstract}
This paper investigates the longitudinal control problem in a dynamic traffic environment where driving scenarios change between free-driving scenarios and car-following scenarios. A comprehensive longitudinal controller is proposed to ensure reasonable transient response and steady-state response in scenarios changes, which is independent of planning algorithms. This design takes into account passenger comfort, safety concerns and disturbance rejections, and attempts to meet the requirement of lower cost, faster response, increased comfort, enhanced safety and elevated extendability from the automated vehicle industry.
Design insights and intuitions are provided in detail. Comprehensive simulations are conducted to demonstrate the efficacy of the proposed controller in different driving scenarios.
\end{abstract}


\section{Introduction}

Last decade witnessed a growing interest in automated vehicles (AVs) due to its potential in enhancing passenger safety, improving travel mobility, reducing fuel consumption, and maximizing traffic throughput \cite{VanderWerfShladover02, Askari_2016, Li_AAP_2017}. Early development starts with research projects in academia, such as DARPA challenges \cite{Campbell_2010}, PATH program \cite{Rajam_Shlad_2001}, Grand Cooperative Driving Challenges \cite{GCDCIntro2011}, etc. Recently the automotive industry is ambitious about equipping production vehicles with automated driving features with higher levels of autonomy \cite{SAE_J3016_2016}. The ultimate objective is to ensure that AVs can drive to destinations autonomously following driving conventions and traffic rules without human attendance and interventions.

The software architecture of AVs mainly include perception, estimation, planning and control. Perception algorithms process raw sensor (camera, radar, lidar, etc.) data, and decipher them to human-readable physical data. Estimation algorithms \cite{Farrell_2017, Wischnewski_2019, Bersani_AEIT_2019} typically apply sensor fusion technique to obtain clean and sound vehicle state estimations based on sensor characteristics. Planning \cite{Karaman_IJRR_2011, Paden_TIV_2016} can be further divided into mission planning (or route planning), behavior planning (or decision making), and motion planning: i) mission planning algorithms select routes to destinations through given road networks based on requirements; ii) behavior planning generates appropriate driving behaviors in real-time according to driving conventions and traffic rules, to guide the interaction with other road-users and infrastructures; iii) motion planning translates the generated behavior primitives into a trajectory based on vehicle states. Control algorithms utilize techniques from control theory \cite{Karl_Richard_Feedback, Khalil_NC, Luenberger_1997, Ioannou_Sun_RAC, Yuri_SMC} that enable vehicle states to follow aforementioned trajectories.

\IEEEpubidadjcol

In longitudinal control, vehicle performance highly depends on motion planning and control algorithms in the standard architecture.
Most longitudinal controllers in literature emphasize on stability, robustness and optimality against steady-state solutions, which are the desired equilibrium states. They rarely consider transient responses in the presence of large initial errors from the equilibria. Thus, these controllers cannot properly handle driving scenario changes in a dynamic traffic environment. Planning algorithms are used as a remedy to solve the problem of transient responses. Planning algorithms generate desired trajectories based on current vehicle state and nominal vehicle models, and then controllers follow the desired vehicle state picked from these trajectories. This strategy decouples the control of transient response and steady-state response into planning and control algorithms, respectively. However, it also leads to issues due to their cohesive relationship. When the nominal models used in planning do not characterize vehicle dynamics well, planning algorithms start to compete against control algorithms due to model mismatches, which might generate unexpected oscillations. Thus, great efforts are spent on the integration and coordination of planning and control algorithms. 

Recently a nonlinear car-following controller inspired by human-driving behaviors was proposed in \cite{Wubing_CF_TVT_2022} that takes passenger comfort and safety into account. It can ensure reasonable transient response and steady-state response while approaching the desired uniform flow equilibrium in car-following scenarios. This motivates us to design longitudinal controllers independent of motion planning algorithms, which can handle both transient response and steady-state response for different driving scenarios. This can avoid the competition between planning outputs based on nominal vehicle models and control outputs based on actual vehicle dynamics. This paper investigates this problem and proposes such a longitudinal controller that is planning-free and computationally cheap. The major contributions as follows. Firstly, the proposed controller integrates the car-following controller proposed in  \cite{Wubing_CF_TVT_2022} for car-following scenarios, and a nonlinear proportional controller for free-driving scenarios. Secondly, a nonlinear integral control is proposed in order to solve the well-known overshooting problem in transient response when driving scenarios change. Moreover, a nonlinear constraint on command changing rate is proposed that can ensure responsiveness in collision-imminent scenario changes to enhance safety, and also avoid jerky motions in normal scenario changes to increase passenger comfort. It is
shown that the proposed controller can ensure reasonable responses similar to human-driving behaviors in dynamic traffic environments when there exist dynamic disturbances and estimation errors.

The layout of this paper is as follows. In Section~\ref{sec:ctrl}, we start with vehicle dynamic modeling and objectives of longitudinal controller design. Then a comprehensive longitudinal controller is proposed that can meet the requirements in a dynamic traffic environment. In Section~\ref{sec:design_details}, we provide details and insights on how the proposed controller is obtained. Simulations are conducted in Section~\ref{sec:res} and results indicate that the proposed controller is effective in scenario changes in the presence of dynamic disturbances and estimation errors. In Section~\ref{sec:conclusion}, we draw conclusions and point out possible future research directions.

\section{Controller Design\label{sec:ctrl}}

In this section we start with the problem statement and present the nonlinear controller.

\subsection{Longitudinal Dynamics \label{sec:extension}}

Based on physics, the longitudinal dynamics of the host vehicle can be modeled as
\begin{equation}\label{eqn:physical_dynamics}
  \begin{split}
    m_{\rm e}\dot{v}_{\rm H} & =\dfrac{\eta\, T}{R} -m g \sin\phi -\mu m g \cos\phi-\rho\, (v_{\rm H}+v_{\rm w})^{2}\,,
  \end{split}
\end{equation}
by neglecting the flexibility of tires. Here, $v_{\rm H}$ is the host vehicle speed, $m_{\rm e}=m+\frac{J}{R^{2}}$ is the effective mass, containing the vehicle static mass $m$, the moment of inertia $J$ of rotating elements, and the wheel radius $R$. Also, $g$ is the gravitational constant, $\phi$ is the inclination angle, $\mu$ is the rolling resistance coefficient, $\rho$ is the air drag constant, $v_{\rm w}$ is the headwind speed, $\eta$ is the gear ratio, and $T$ is the actuation torque. The actuation torque $T$ is also governed by actuator dynamics, which is typically modeled as a first-order system, that is,
\begin{equation}\label{eqn:actuator_dyn}
  \dot{T} = -\dfrac{1}{\tau}\, T +\dfrac{1}{\tau}\, T_{\rm des}\,,
\end{equation}
where $T_{\rm des}$ is the desired torque, and $\tau$ is the time constant.

In practice automated vehicles typically follow a hierarchical design architecture that consists of a high-level longitudinal controller and a low-level actuator controller \cite{He_2019}. These high-level controllers generate acceleration command $u$ that can achieve normal cruise control in free-driving scenario, or adaptive cruise control in car-following scenario. Low-level controllers generate desired actuation torque $T_{\rm des}$ based on acceleration command $u$ such that vehicle acceleration $\dot{v}_{\rm H}$ can follow the given acceleration command $u$. The main focus of this paper is on the high-level longitudinal controller design. Thus, in the following we take a simple low-level controller design as an example. One can apply other techniques on low-level controller design as well.

Applying feedback linearization technique on model \eqref{eqn:physical_dynamics}, one can: i) set $\dot{v}_{\rm H}$ to the given acceleration command $u$; ii) set vehicle static mass $m$, rolling resistance coefficient $\mu$ and air drag constant $\rho$ to their nominal values $\hat{m}$, $\hat{\mu}$ and $\hat{\rho}$, respectively; and iii) set the inclination angle $\phi$ to the estimated value $\hat{\phi}$ that can be obtained through onboard sensors (IMU+wheel-based acceleration measurements). Then the solution of actuation torque $T$ is the desired value $T_{\rm des}$ based on acceleration command $u$ and other parameters, which yields the low-level controller
\begin{equation}\label{eqn:low_level_trq_ctrl}
T_{\rm des} =\dfrac{R}{\eta}(\hat{m}\:\! u+\hat{m} g \sin\hat{\phi}+\hat{\mu} \hat{m} g\cos\hat{\phi}+\hat{\rho}\, v_{\rm H}^{2})\,.
\end{equation}
To obtain the backbone longitudinal dynamics, we define the host vehicle acceleration as
\begin{equation}\label{eqn:follower_accel}
  a_{\rm H} = \dot{v}_{\rm H}\,,
\end{equation}
and derive its derivative to transform the state variable from actuator torque $T$ to acceleration $a_{\rm H}$.
Differentiating model \eqref{eqn:physical_dynamics} and utilizing (\ref{eqn:physical_dynamics}-\ref{eqn:follower_accel}) in the substitution, we obtain
\begin{equation}\label{eqn:long_dynamics}
  \begin{split}
    \dot{v}_{\rm H} & = a_{\rm H}\,,\\
    \dot{a}_{\rm H} &= -\dfrac{1}{\tau}\, a_{\rm H} +\dfrac{\alpha_{1}}{\tau}\, u +\dfrac{\alpha_{1}}{\tau}\,\Delta\,.
  \end{split}
\end{equation}
by choosing the host vehicle speed $v_{\rm H}$ and acceleration $a_{\rm H}$ as state variables, where the disturbance is
\begin{equation}
\begin{split}
\Delta &=\big(\sin\hat{\phi}+\hat{\mu}\cos\hat{\phi}-\alpha_{2}(\sin\phi+\mu\cos\phi)\big)\,g\\
 &+\alpha_{2}\,g\,\tau\,\dot{\phi}\,(\mu\sin\phi-\cos\phi)\\
 &+\dfrac{{\hat{\rho}\,v_{\rm H}^{2}-\rho\, (v_{\rm H}+v_{\rm w})^{2}}
 -{2\rho\,\tau}(v_{\rm H}+v_{\rm w})\,(a_{\rm H}+\dot{v}_{\rm w})}{\hat{m}}\,,
\end{split}
\end{equation}
and
\begin{align}\label{eqn:ratio_alpha}
  \alpha_{1} &=\dfrac{\hat{m}}{m_{\rm e}}\,,&
  \alpha_{2} &=\dfrac{m}{\hat{m}}\,.
\end{align}
Model \eqref{eqn:long_dynamics} characterizes the backbone longitudinal dynamics of a single vehicle, which can be utilized in normal cruise control design. We remark that the disturbance $\Delta$ lumps the errors between nominal/estimated values and actual values, and the effects of headwind together. When there is no heavy headwind, this disturbance is typically small and varies in a random manner due to changes in road grade, road roughness, air temperature, etc. In the presence of heavy headwind, it varies around a noticeable offset due to the ignorance of headwind.

\begin{figure}
  \centering
  \includegraphics[scale=0.6]{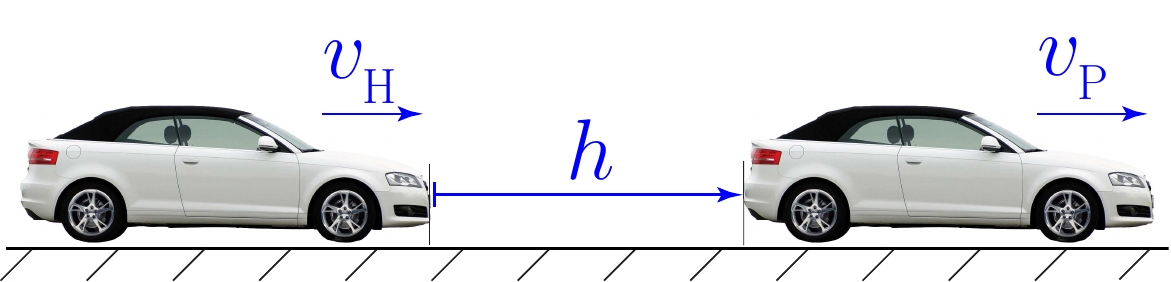}\\
  \caption{Car-following scenario.}\label{fig:car_follow}
\end{figure}

To characterize the car-following scenario shown in Fig.~\ref{fig:car_follow}, the dynamics on relative motion between the host vehicle and the preceding vehicle need to be appended to the single vehicle dynamics \eqref{eqn:long_dynamics}, leading to
\begin{equation}\label{eqn:car_following_dyn}
  \begin{split}
    \dot{h} & = v_{\rm P} -v_{\rm H}\,,\\
    \dot{v}_{\rm H} & = a_{\rm H}\,,\\
    \dot{a}_{\rm H} &= -\dfrac{1}{\tau}\, a_{\rm H} +\dfrac{\alpha_{1}}{\tau}\, u +\dfrac{\alpha_{1}}{\tau}\,\Delta\,,
  \end{split}
\end{equation}
where $v_{\rm P}$ and $h$ represent the preceding vehicle speed and the inter-vehicle distance, respectively.

\subsection{Objectives \label{sec:objectives}}

We assume that the host vehicle can obtain: i) its own speed $v_{\rm H}$ with onboard sensors; and ii) the preceding vehicle speed $v_{\rm P}$ and inter-vehicle distance $h$ via onboard sensors or V2V communication in car-following scenario.
The objective is to design a comprehensive longitudinal controller for the host vehicle in a dynamic traffic environment that meets the following objectives:
\begin{enumerate}
  \item In \emph{free-driving }scenario where there exists no preceding vehicles, the host vehicle switches to normal \emph{cruise control} with preset maximum speed $v_{\max}$.
  \item In \emph{car-following} scenario where there exist preceding vehicles, the host vehicle adapts its speed $v_{\rm H}$ and inter-vehicle distance $h$ according to the motion of the preceding vehicle, which is referred to as \emph{adaptive cruise control}. When the preceding vehicle speed $v_{\rm P}$ is constant, the host vehicle must be able to match its speed $v_{\rm H}$ with the preceding vehicle speed $v_{\rm P}$ while maintaining the desired distance $h_{\rm des}$ given by the so-called \emph{range policy}. This requires the possession and stabilizability of so-called \emph{uniform flow equilibrium}.
  \item In case of emergency within physical capability, the host vehicle is able to avoid collision.
  \item When the driving scenario changes, the host vehicle must respond reasonably in the transient phase while approaching the equilibrium (cruise control with $v_{\max}$ in free-driving scenario, or uniform flow in car-following scenario). These scenario changes include changes in preceding vehicles and changes in the existence of preceding vehicles that are caused by cut-in, cut-out maneuvers in a dynamic traffic environment.
  \item The host vehicle must be able to adapt to variations in dynamic disturbances $\Delta$.
\end{enumerate}
Regarding the range policy, we use the constant time-headway policy in this paper, i.e.,
\begin{align}\label{eqn:hdes}
  h_{\rm des} &= h_{0} + v_{\rm P}\, t_{\rm h} \,,
\end{align}
where $h_{0}$ is the standstill distance, and $t_{\rm h}$ is the desired time-headway. We remark that this desired distance is based on the predecessor speed instead of the follower speed. One can refer to \cite{Wubing_CF_TVT_2022} for more details on this design.

\subsection{Longitudinal Controller Design \label{sec:long_ctrl_design}}

We propose the following controller
\begin{subequations}\label{eqn:ctrl_nonlinear_PI_rate_lim}
  \begin{align}
    \dot{u} &= r_{\max}\, g\big(\tfrac{k_{\rm u}}{ r_{\max}}(u_{\rm des}-u)\big)\,,\label{eqn:ctrl_nonlinear_rate_lim}\\
    u_{\rm des} & =a_{\rm des} + k_{\rm i}\, e\,, \label{eqn:ctrl_udes}\\
    \dot{e} & = \sigma\; p\big(\tfrac{v_{\rm des}-v_{\rm H}}{\sigma}\big)\,,\label{eqn:ctrl_nonlinear_integrator}
  \end{align}
\end{subequations}
where $r_{\max}$ is the maximum allowed changing rate of command acceleration, $k_{\rm u}$ and $k_{\rm i}$ are control gains, $u_{\rm des}$ is the desired command acceleration, $\sigma$ is a parameter in $[\frac{\rm m}{\rm s}]$ characterizing the effective speed range of the integrator, and $g(\cdot)$ and $p(\cdot)$ are shaping functions. The desired acceleration $a_{\rm des}$ and desired velocity $v_{\rm des}$ will be designed separately in free-driving and car-following scenarios. We remark that (\ref{eqn:ctrl_nonlinear_integrator}) is a nonlinear integral controller to ensure adaptation to variations in disturbance $\Delta$, while \eqref{eqn:ctrl_nonlinear_rate_lim} sets constraint on the changing rate to ensure smooth transition when driving scenarios change. The desired acceleration $a_{\rm des}$ in \eqref{eqn:ctrl_udes} will contain a nonlinear proportional controller to ensure stabilizability of the desired equilibria in different driving scenarios. In summary, controller \eqref{eqn:ctrl_nonlinear_PI_rate_lim} is a nonlinear proportional-integral controller with nonlinear rate limiting constraints to fit dynamic traffic environment.  More design details will be provided later.

\begin{figure}
  \centering
  \includegraphics[scale=1.1]{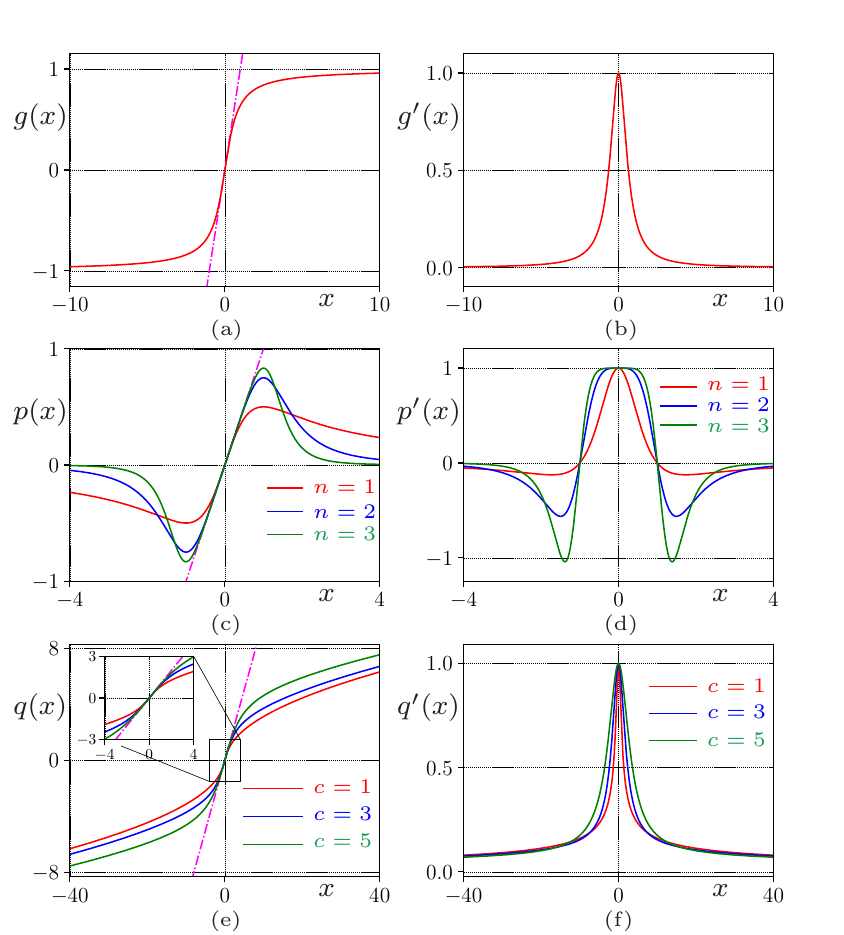}\\
  \caption{(a, b) Function $g(x)$ and its derivative $g'(x)$. (c, d) Function $p(x)$ and its derivative $p'(x)$ for $n=1,\,2,\,3$. (e, f) Function $q(x)$ and its derivative $q'(x)$ for different $c$ value when $b=0.5$.}\label{fig:wrappers}
\end{figure}

Function ${g:\, \mathbb{R}\to \mathbb{R} }$ denotes a shaping function satisfying the following properties:
\begin{enumerate}
  \item it is odd over $\mathbb{R}$, i.e., ${g(x)=-g(-x)}$ for ${x\in \mathbb{R}}$;
  \item it is continuously differentiable and strictly increasing.;
  \item it is bounded in $[-1, \, 1]$, i.e., $g:\mathbb{R}\to [-1, \, 1]$;
  \item its derivative strictly decreases over ${\mathbb{R}_{\geq0}}$ such that ${g'(0)=1}$ and ${\lim\limits_{x \to +\infty}  g'(x) = 0}$.
\end{enumerate}
Also, function ${p:\, \mathbb{R}\to \mathbb{R} }$ denotes another shaping function satisfying the following properties:
\begin{enumerate}
  \item it is odd over $\mathbb{R}$, i.e., ${p(x)=-p(-x)}$ for ${x\in \mathbb{R}}$;
  \item it is continuously differentiable over $\mathbb{R}$;
  \item it is positive over ${\mathbb{R}_{> 0}}$ and ${\lim\limits_{x \to +\infty}  p(x) = 0}$;
  \item it is strictly increasing over $[0, 1]$ and strictly decreasing over $[1, +\infty)$, implying $p'(1)=0$;
  \item its derivative is continuous over ${\mathbb{R}}$ and $p'(0)=1$.
\end{enumerate}
In this paper we use the smooth wrapper function
\begin{subequations}
  \begin{align}
    g(x) &=\tfrac{2}{\pi}\arctan \big(\tfrac{\pi}{2} x\big)\,, \label{eqn:func_g}\\
    p(x) &= \dfrac{x}{1+\frac{1}{2n-1}x^{2n}}\,, & n=1,\, 2,\,3,\,\ldots \label{eqn:func_p}
  \end{align}
\end{subequations}
In Fig.~\ref{fig:wrappers}, panels (a, b) plot the function $g(x)$ and its derivative $g'(x)$, while panels (c, d) plot the function $p(x)$ and its derivative $p'(x)$ for $n=1,\, 2,\,3$.

\subsubsection{Free-driving Scenario}
When there exists no preceding vehicles, host vehicle is in normal cruise control mode, and the objective is to reach and maintain the preset maximum speed $v_{\max}$. Thus, we set
\begin{equation}\label{eqn:vdes_CC}
  v_{\rm des} = v_{\max}\,,
\end{equation}
and propose a nonlinear proportional controller
\begin{equation}\label{eqn:ades_CC}
  a_{\rm des} = a_{\sat}\; g\big(\tfrac{k_{\rm v}\, (v_{\rm des}-v_{\rm H})}{a_{\sat}}\big)\,,
\end{equation}
which is used to ensure that the host vehicle speed $v_{\rm H}$ can reach and maintain the desired speed $v_{\max}$. Here, $a_{\sat}>0$ is the maximum allowed acceleration, $k_{\rm v}$ is a control gain, and $g(\cdot)$ is the aforementioned shaping function.

\subsubsection{Car-following Scenario}
When there is a preceding vehicle, the host vehicle is in adaptive cruise control mode and must be able to avoid collisions. The nonlinear car-following controller proposed in \cite{Wubing_CF_TVT_2022} is capable of avoiding collision and performs reasonably in both the transient response and steady-state response while approaching and maintaining the uniform flow equilibrium. Thus, we integrate this car-following controller here and refer readers to \cite{Wubing_CF_TVT_2022} for more details. By defining the errors as
\begin{align}\label{eqn:delta_v_h}
  \hat{v} &=v_{\rm P}-v_{\rm H}\,, &
  \hat{h} & = h - h_{\rm des}\,,
\end{align}
cf.~\eqref{eqn:hdes}, we design the desired speed
\begin{align}\label{eqn:vdes_acc}
  v_{\rm des} &= \max\bigg\{\min\Big\{v_{\rm P} + q\big(k_{\rm h}\, \hat{h};\, \tfrac{a_{\rm com}}{k_{\rm h}}\big),\, v_{\max} \Big\}, \,0\bigg\}\,,
\end{align}
and the desired acceleration
\begin{align}\label{eqn:ades_acc}
  a_{\rm des} & = a_{\sat}\; g\big(\tfrac{k_{\rm v}\, (v_{\rm des}-v_{\rm H})}{a_{\sat}}\big)+\bar{a}_{\rm fb}+a_{\rm cf}\,,
\end{align}
where the same notations are maintained as those in (\ref{eqn:vdes_CC}, \ref{eqn:ades_CC}). Also, $k_{\rm h}$ is another control gain, $a_{\rm com}$ is the comfortable acceleration applied in the transient phase. The collision-free feedforward acceleration $a_{\rm cf}$ and the acceleration term $\bar{a}_{\rm fb}$ will be discussed later. Function $q(x; \, b)$ represents another shaping function with a parameter $b>0$ satisfying the following properties:
\begin{enumerate}
  \item it is continuously differentiable and strictly increasing over $\mathbb{R}$;
  \item it is odd, i.e., ${q(x)=-q(-x)}$ for ${x\in \mathbb{R}}$;
  \item it has a curvilinear asymptote $y=\sqrt{2\,b\, x}$ as $x\to +\infty$, i.e., $ \lim\limits_{x \to +\infty}  \left\{q(x)-\sqrt{2\,b\, x}\right\} = 0$;
  \item its derivative is continuous and strictly decreasing over ${\mathbb{R}_{\geq0}}$ and $q'(0)=1$.
\end{enumerate}
Note that here we use the shorthand notation $q(x)$ to represent $q(x;\,b)$ when highlighting parameters is not necessary. This notation will be maintained for other functions as well throughout the paper.
In this paper, we use the smooth shaping function
\begin{align}\label{eqn:wrapper_q}
  q(x; b) &= g(\tfrac{x}{c})\sqrt{2\,b\, x\, g(\tfrac{x}{c})+c^{2}}\,,
\end{align}
where $c>0$ is a slackness parameter and $g(x)$ is the wrapper function \eqref{eqn:func_g}. Fig.~\ref{fig:wrappers}(e, f) shows function $q(x)$ and its derivative.

As explained in \cite{Wubing_CF_TVT_2022}, the desired speed \eqref{eqn:vdes_acc} is inspired by human-driving behaviors that tend to approach the desired uniform flow equilibrium with a near-constant comfortable acceleration $a_{\rm com}$. We highlight that this desired speed depends on the preceding vehicle speed $v_{\rm P}$ and the inter-vehicle distance $h$, which is demonstrated to ensure reasonable transient behaviors and steady-state behaviors. Regarding the desired acceleration $a_{\rm des}$ in \eqref{eqn:ades_acc}, the first term  is a nonlinear feedback control to ensure that the host vehicle speed $v_{\rm H}$ follows the desired speed $v_{\rm des}$, while the second term $\bar{a}_{\rm fb}$ is the underlying acceleration that is required when $v_{\rm H}$ tracks $v_{\rm des}$ perfectly. Differentiating \eqref{eqn:vdes_acc}, we obtain the underlying acceleration
\begin{equation}\label{eqn:fb_ades_S}
  \bar{a}_{\rm fb} =
  \left\{
    \begin{array}{ll}
      \max\Big\{q'\big(k_{\rm h}\, \hat{h};\tfrac{a_{\rm com}}{k_{\rm h}}\big)k_{\rm h}\hat{v}\,,\; 0\Big\}
            \enskip \hbox{if $v_{\rm des}=0$,} \vspace{1mm} \\
      q'\big(k_{\rm h}\, \hat{h};\tfrac{a_{\rm com}}{k_{\rm h}}\big)k_{\rm h}\hat{v}
            \qquad \hbox{if $0<v_{\rm des}<v_{\max}$,} \vspace{1mm} \\
      \min\Big\{q'\big(k_{\rm h}\, \hat{h};\tfrac{a_{\rm com}}{k_{\rm h}}\big)k_{\rm h}\hat{v}\,,\; 0\Big\}
            \enskip \hbox{otherwise,}
    \end{array}
  \right.
\end{equation}
where $v_{\rm P}$ is considered constant. Note that when $v_{\rm des}$ reaches saturation limits, \eqref{eqn:vdes_acc} is not differentiable. In these special cases, the underlying acceleration $\bar{a}_{\rm fb}$ is designed such that it is continuous with respect to $v_{\rm des}$, and can also avoid unexpected acceleration or deceleration when $v_{\rm des}$ saturates to $v_{\max}$ or $0$.

To avoid collisions, the last term of desired acceleration $a_{\rm des}$ in \eqref{eqn:ades_acc} is the collision-free feedforward law
\begin{align}\label{eqn:ff_cf_term}
  a_{\rm cf} &=  \max\left\{-\dfrac{\hat{v}^{2}\cdot H(-\hat{v})}{2\,\max\{h-h_{\min},\, \varepsilon\} }\,,\; a_{\min} \right\}\,,
\end{align}
where $H(x)$ is the heaviside step function, $h_{\min}$ is the minimum allowed inter-vehicle distance, $\varepsilon>0$ is used to avoid singularity, and $a_{\min}<0$ is the physical minimum acceleration.


\section{Design Details \label{sec:design_details}}
In this section, we provide design details on the controller. Readers who are not interested in the details, may jump to Section~\ref{sec:res} for simulation results.

In a dynamic traffic environment, it is common that driving scenarios may change in the following way:
\begin{itemize}
  \item the current preceding vehicle cuts out such that car-following scenario changes to free-driving scenario;
  \item a vehicle in the adjacent lane cuts in such that free-driving scenario changes to car-following scenario;
  \item a vehicle in the adjacent lane cuts in behind the current preceding vehicle such that car-following scenario changes the preceding vehicle ($v_{\rm P}$ and $h$ are changed).
\end{itemize}
Most longitudinal controllers in literature emphasize on stability, robustness and optimality against desired equilibria. They rarely consider transient responses in the presence of large initial errors from the equilibria, which is very common in a dynamic traffic environment. When driving scenarios change, control mode, preceding vehicle speed $v_{\rm P}$ and inter-vehicle distance $h$ may change abruptly and host vehicle states can be far away from the desired equilibria in case of new driving mode or a new preceding vehicle. Thus, these controllers cannot properly handle such scenario changes, and planning algorithms are used as a remedy to solve the problem of transient responses while approaching the desired equilibria.  Planning algorithms generate desired trajectories based on current vehicle state and nominal vehicle models, and then controllers follow the desired vehicle state picked from these trajectories. This strategy is widely used since it decouples the control of transient response and steady-state response into planning and control algorithms. However, it also leads to issues due to their cohesive relationship. When the nominal model used in planning does not characterize vehicle dynamics well, planning algorithms start to compete against control algorithms due to model mismatches, which might generate unexpected oscillations. Thus, great efforts are spent on the integration and coordination of planning and control algorithms.

Tasks for planning might be reduced if control algorithms can handle both transient response and steady-state response reasonably. A nonlinear car-following controller is proposed in \cite{Wubing_CF_TVT_2022} that can ensure reasonable transient response while approaching the steady-state equilibria when following a preceding vehicle. The performance is similar to human-driving behaviors when preceding vehicle changes. Thus, we might come up with a comprehensive longitudinal controller if we can: i) design a cruise controller for free-driving scenario that can also ensure reasonable transient response, ii) integrate the cruise controller with the car-following controller, and iii) design a mechanism to handle the instant when driving scenarios change. We follow these guidelines and propose the controller provided in Section~\ref{sec:ctrl}. In the following, more details are provided.

\subsection{Constraint on Command Changing Rate}

Scenario changes lead to switching in vehicle control, either between normal cruise control and adaptive cruise control, or adaptation to a different preceding vehicle. Thus, it is inevitable that sudden changes will appear in desired acceleration $a_{\rm des}$ (cf.~(\ref{eqn:vdes_CC},\ref{eqn:ades_CC}) and (\ref{eqn:vdes_acc},\ref{eqn:ades_acc})) and desired command acceleration $u_{\rm des}$ (cf.~\eqref{eqn:ctrl_udes}). Applying this desired command acceleration $u_{\rm des}$ directly to the actuator, i.e., $u=u_{\rm des}$, will lead to abrupt changes in actuator control (cf.~\eqref{eqn:low_level_trq_ctrl}). Consequently, jerky motion, noises and vibrations may appear which deteriorate passenger comfort. Thus, it is common in practice to add a constraint on the command changing rate to avoid abrupt changes.

The simple strategy is using constant maximum rate $r_{\max}$ to limit abrupt changes, which is commonly-used in practice. This can be formulated as
\begin{equation}\label{eqn:const_rate}
  \begin{split}
    \dot{u} &= r_{\max}\, \sign{u_{\rm des}-u}\,,\\
  \end{split}
\end{equation}
which applies a constant rate $r_{\max}$ (or  $-r_{\max}$) to transit from current command to desired command $u_{\rm des}$ when there is an increase (or decrease). However, it is controversial to set this constant rate $r_{\max}$ because: i) $r_{\max}$ should be large enough to ensure responsiveness in case of emergency; and ii) large values of $r_{\max}$ lead to noticeable jerky motions in case of moderate changes in desired command $u_{\rm des}$. To resolve this issue, we change the sign function in \eqref{eqn:const_rate} to a smoother shaping function $g(\cdot)$ in \eqref{eqn:ctrl_nonlinear_rate_lim} such that the rate is still constrained in ${[-r_{max}\,, r_{\max}]}$ and can change monotonically with respect to the magnitude of difference between the current command and desired command, i.e., $|u_{\rm des}-u|$. When this difference is large, the changing rate in \eqref{eqn:ctrl_nonlinear_rate_lim} is saturated and can be approximated by \eqref{eqn:const_rate}. While the difference is small, the changing rate in \eqref{eqn:ctrl_nonlinear_rate_lim} is in the linear range and can be approximate by
\begin{equation}\label{eqn:linear_rate}
  \begin{split}
    \dot{u} &= k_{\rm u}\, (u_{\rm des}-u)\,.\\
  \end{split}
\end{equation}
Note that when

\subsection{Nonlinear Integral Control}

When there exists unknown disturbances $\Delta$ in vehicle dynamics, it is common and effective to utilize integral control technique to ensure the existence of the desired equilibria. This is because accumulation of integration errors will in turn update control commands to eliminate errors caused by disturbances.  To ensure that host vehicle speed $v_{\rm H}$ eventually equates desired speed $v_{\rm des}$, the standard integral control is
\begin{equation}\label{eqn:ctrl_int_normal}
  \begin{split}
    \dot{e} & = v_{\rm des}-v_{\rm H}\,,
  \end{split}
\end{equation}
which is the linearized version of \eqref{eqn:ctrl_nonlinear_integrator} when the difference $v_{\rm des}-v_{\rm H}$ is relatively small.
This integral control \eqref{eqn:ctrl_int_normal} is very effective to ensure steady state equilibrium in the presence of disturbance. However, its underlying issue in transient response is also famous. When initial errors $v_{\rm des}-v_{\rm H}$ (in scenario changes) are large, error $e$ accumulates fast and always leads to oscillations caused by overshoots/undershoots. The overshoot percentage has a positive correlation with initial errors. When driving scenarios change, it is quite common that the difference $v_{\rm des}-v_{\rm H}$ is large. Thus, integral controller \eqref{eqn:ctrl_int_normal} typically generates unexpected overshoots/undershoots while switching to new driving scenarios.

We attempt to resolve this issue with the nonlinear integral controller \eqref{eqn:ctrl_nonlinear_integrator}. The properties of $p(x)$ indicate that $p(x)=0$ is equivalent to $x=0$, implying that the steady-state equilibrium remains unchanged. That is, host vehicle speed $v_{\rm H}$ is equal to desired speed $v_{\rm des}$. Also, properties of $p(x)$ implies that the controller \eqref{eqn:ctrl_nonlinear_integrator} is topologically equivalent to the controller \eqref{eqn:ctrl_int_normal} when the difference $v_{\rm des}-v_{\rm H}$ is small enough. Thus, the performance of standard integral controller \eqref{eqn:ctrl_int_normal} is preserved in such scenarios. Moreover, properties of $p(x)$ indicates that integration of large difference in $v_{\rm des}-v_{\rm H}$ is inhibited because the function value $p(x)$ is significantly suppressed for large $|x|$ when $|x|\gg 1$. Hence, parameter $\sigma$ in controller \eqref{eqn:ctrl_nonlinear_integrator} can be viewed as the effective range of the integral control, and integration of difference with magnitude larger than $\sigma$ are suppressed. Fig.~\ref{fig:wrappers}(c, d) plot the shaping function \eqref{eqn:func_p} and its derivative when $n=1$, $2$ and $3$. One can see that the higher the order $n$ is, the faster the function suppresses the value for large $|x|$.

\subsection{Nonlinear Proportional Control}

Nonlinear proportional control technique is applied in controller \eqref{eqn:ades_CC} and controller \eqref{eqn:ades_acc} for free-driving and car-following scenarios, respectively. The design details for controller \eqref{eqn:ades_acc} are provided in \cite{Wubing_CF_TVT_2022}. Thus, in this part we explain the details of controller \eqref{eqn:ades_CC} for free-driving scenario.

When there exists no preceding vehicles, host vehicle starts cruise control mode. The desired speed $v_{\rm des}$ is the driver preset maximum speed $v_{\max}$, and the desired acceleration $a_{\rm des}$ may apply a feedback law to ensure that host vehicle speed $v_{\rm H}$ can reach the desired speed $v_{\rm des}$ when there is no disturbance. A standard linear feedback law is
\begin{equation}\label{eqn:ades_CC_lin}
  a_{\rm des} = k_{\rm v}\, (v_{\rm des}-v_{\rm H})\,,
\end{equation}
where $k_{\rm v}>0$ is the feedback gain.
Note that this feedback law may saturate to the maximum allowed acceleration $a_{\rm sat}>0$ when the error $v_{\rm des}-v_{\rm H}$ is too large, which can be formulated as
\begin{equation}\label{eqn:ades_CC_lin_sat}
  a_{\rm des} = \min\Big\{\max\big\{k_{\rm v}\, (v_{\rm des}-v_{\rm H}), -a_{\rm sat}\big\}, a_{\rm sat}\Big\}\,.
\end{equation}
Controller \eqref{eqn:ades_CC_lin_sat} applies a constant gain $k_{\rm v}$ in the linear range. In practice it is favorable to utilize large gains such that host vehicle speed $v_{\rm H}$ tracks the desired speed $v_{\rm des}$ well, because in such case small errors in $v_{\rm des}-v_{\rm H}$ may lead to noticeable change in $a_{\rm des}$. However, large gains cause ``overreaction" issue because a moderate difference in $v_{\rm des}-v_{\rm H}$ may result in saturations. To resolve the issue, we update controller \eqref{eqn:ades_CC_lin_sat} into controller \eqref{eqn:ades_CC} using a shaping function \eqref{eqn:func_g}. Fig.~\ref{fig:wrappers}(a) indicates that the desired acceleration is still bounded in $[ -a_{\rm sat},\, -a_{\rm sat}]$. Fig.~\ref{fig:wrappers}(b) implies that the effective gain decreases monotonically from $k_{\rm v}$ to $0$ as the magnitude of $|v_{\rm des}-v_{\rm H}|$ increase, because the derivative of function $g(x)$ decreases monotonically from $1$ to $0$ as $|x|$ increases.

\subsection{Steady-state in Free-driving Scenario}



In free-driving scenario, host vehicle is in cruise control mode. It can be seen that the closed-loop dynamics (\ref{eqn:long_dynamics},\ref{eqn:ctrl_nonlinear_PI_rate_lim},\ref{eqn:vdes_CC},\ref{eqn:ades_CC}) possesses the desired equilibrium
\begin{align}\label{eqn:equil_CC}
  v_{\rm H}^{\ast} &= v_{\max}\,,&
  a_{\rm H}^{\ast} &=0\,,&
  u^{\ast} &=-\Delta^{\ast}\,,&
  e^{\ast} &=-\tfrac{\Delta^{\ast}}{k_{\rm i}}\,.
\end{align}
Defining the perturbations as
\begin{equation}\label{eqn:perturb_CC}
  \begin{split}
    \tilde{v}_{\rm H} &= v_{\rm H}-v_{\rm H}^{\ast}\,,\quad
    \tilde{a}_{\rm H} =a_{\rm H}-a_{\rm H}^{\ast}\,, \quad
    \tilde{u} = u- u^{\ast}\,, \\
    \tilde{e} &= e-e^{\ast}\,,\quad\quad\;
    \tilde{\Delta} = \Delta-\Delta^{\ast}
  \end{split}
\end{equation}
we obtain the linearized dynamics
\begin{equation}\label{eqn:lin_dynamics_CC}
  \dot{\bfx}_{1} = \bfA_{1}\, \bfx_{1} +\bfB_{1}\,\bfu_{1}\,,
\end{equation}
where the state and input are
\begin{align}\label{eqn:lin_dyn_CC_x_u}
  \bfx_{1} &=
    \begin{bmatrix}
        \tilde{v}_{\rm H} & \tilde{a}_{\rm H} &  \tilde{u} & \tilde{e}
    \end{bmatrix}^{\top}\,,&
  \bfu_{1} &= \tilde{\Delta}\,,
\end{align}
and matrices are
\begin{align}\label{eqn:lin_dyn_CC_AB}
  \bfA_{1} &=
    \begin{bmatrix}
        0 & 1 & 0 & 0\\
        0 &  -\tfrac{1}{\tau} & \tfrac{\alpha_{1}}{\tau} & 0\\
        -k_{\rm u} k_{\rm v} & 0 & -k_{\rm u} & k_{\rm u} k_{\rm i}\\
        -1 & 0 & 0 & 0
    \end{bmatrix}, \quad
    \bfB_{1} =
    \begin{bmatrix}
        0 \\ \tfrac{\alpha_{1}}{\tau} \\ 0 \\ 0
    \end{bmatrix}\,.
\end{align}
The characteristic equation is
\begin{equation}
  \begin{split}
    \det(s\, \bfI&-\bfA_{1})=
    s^4+(k_{\rm u}+\tfrac{1}{\tau})\, s^3 +\tfrac{k_{\rm u}}{\tau} s^2 \\
    &+\tfrac{ \alpha_{1}}{\tau}k_{\rm u} k_{\rm v}\, s +\tfrac{ \alpha_{1}}{\tau} k_{\rm u} k_{\rm i}\,,
  \end{split}
\end{equation}
and one can apply Routh-Hurwitz criterion to derive stability conditions.

\subsection{Steady-state in Car-following Scenario}

%

In car-following scenario, host vehicle is in adaptive cruise control mode. It can be seen that the closed-loop dynamics (\ref{eqn:car_following_dyn},\ref{eqn:ctrl_nonlinear_PI_rate_lim},\ref{eqn:delta_v_h}-\ref{eqn:ff_cf_term}) possesses the desired uniform flow equilibrium
\begin{equation}\label{eqn:equil_ACC}
  \begin{split}
  h^{\ast} &= h_{0}+t_{\rm h}\,v_{\rm P}^{\ast}\,, \qquad
  v_{\rm H}^{\ast} = v_{\rm P}^{\ast}\,,\\
  a_{\rm H}^{\ast} &=0\,,\qquad
  u^{\ast} =-\Delta^{\ast}\,,\qquad
  e^{\ast} =-\tfrac{\Delta^{\ast}}{k_{\rm i}}\,.
\end{split}
\end{equation}
Defining the same perturbations as \eqref{eqn:perturb_CC} and
\begin{equation}\label{eqn:perturb_ACC}
  \begin{split}
    \tilde{v}_{\rm P} &= v_{\rm P}-v_{\rm P}^{\ast}\,,\qquad
    \tilde{h} = h- h^{\ast}\,,
  \end{split}
\end{equation}
we obtain the linearized dynamics
\begin{equation}\label{eqn:lin_dynamics_ACC}
  \dot{\bfx}_{2} = \bfA_{2}\, \bfx_{2} +\bfB_{2}\,\bfu_{2}\,,
\end{equation}
where the state and input are
\begin{align}\label{eqn:lin_dyn_ACC_x_u}
  \bfx_{2} &=
    \begin{bmatrix}
        \tilde{h} & \tilde{v}_{\rm H} & \tilde{a}_{\rm H} &  \tilde{u} & \tilde{e}
    \end{bmatrix}^{\top}\,, &
  \bfu_{2} &=
    \begin{bmatrix}
        \tilde{\Delta} & \tilde{v}_{\rm P}
    \end{bmatrix}^{\top}\,,
\end{align}
and the matrices are
\begin{equation}
\begin{split}\label{eqn:lin_dyn_ACC_A}
  \bfA_{2} &=
    \begin{bmatrix}
        0 & -1 & 0 & 0 & 0\\
        0 & 0 & 1 & 0 & 0\\
        0 & 0 &  -\tfrac{1}{\tau} & \tfrac{\alpha_{1}}{\tau} & 0\\
        k_{\rm u} k_{\rm v} k_{\rm h} & -k_{\rm u} (k_{\rm v}+ k_{\rm h}) & 0 & -k_{\rm u} & k_{\rm u} k_{\rm i}\\
        k_{\rm h} & -1 & 0 & 0 & 0
    \end{bmatrix}\,, \\
  \bfB_{2} &=
    \begin{bmatrix}
        0 & 1\\
        0 & 0\\
        \tfrac{\alpha_{1}}{\tau} & 0\\
        0 & k_{\rm u} (k_{\rm v}+ k_{\rm h})\\
        0 & 1
    \end{bmatrix}\,.
\end{split}
\end{equation}
The characteristic equation is
\begin{equation}
  \begin{split}
    \det(s\, \bfI&-\bfA_{2})=
    s^5+(k_{\rm u}+\tfrac{1}{\tau})\, s^4 +\tfrac{k_{\rm u}}{\tau} s^3 +\tfrac{ \alpha_{1}}{\tau}k_{\rm u}(k_{\rm h}+ k_{\rm v})\, s^2\\
    &+\tfrac{ \alpha_{1}}{\tau}k_{\rm u}(k_{\rm i}+k_{\rm v} k_{\rm h})\, s+\tfrac{ \alpha_{1}}{\tau}k_{\rm h} k_{\rm u} k_{\rm i}\,,
  \end{split}
\end{equation}
and one can also derive stability conditions on the parameters.

%

\section{Results \label{sec:res}}

\begin{table}[!t]
\begin{center}
\renewcommand{\arraystretch}{1.3}
\rowcolors{1}{Azure}{LightGreen}
\begin{tabular}{l|c|l}
\hline\hline
 \rowcolor{Gray}  Parameter & Value & Description\\
 \hline 
 $h_{0}$ [m]& $5$ & standstill distance\\
 $t_{\rm h}$ [s]& $1$ & desired time headway\\
 $h_{\rm min}$ [m]& $5$ & minimum allowed distance\\
 $\varepsilon$ [m]& $0.5$ & small value to avoid singularity\\
 $v_{\max}$ [m/s]& $30$ & driver preset maximum speed\\
 $r_{\max}$ [m/s$^{3}$] & $5$ & maximum rate of acceleration command\\
 $a_{\sat}$ [m/s$^{2}$] & $4$ & maximum allowed acceleration\\
 $a_{\min}$ [m/s$^{2}$] & $-10$ & physical minimum acceleration\\
 $a_{\rm com}$ [m/s$^{2}$] & $0.5$ & user-specific comfortable acceleration\\
 $k_{\rm v}$ [s$^{-1}$]& $0.8$ & control gain\\
 $k_{\rm h}$ [s$^{-1}$]& $1$ & control gain\\
 $k_{\rm i}$ [s$^{-1}$]& $0.08$ & control gain\\
 $k_{\rm u}$ [s$^{-1}$]& $10$ & control gain\\
 $c$ [m/s]& $0.5$ & slackness parameter in function $q(x)$\\
 $n$ [1]& $2$ & order of function $p(x)$ in \eqref{eqn:func_p}\\
 $\sigma$ [m/s]& $1$ & effective range of integral controller\\
 $\tau$ [s] & $0.5$ & time constant of actuator dynamics\\
 $\alpha_{1}$ [1] & $1$ & ratio in \eqref{eqn:ratio_alpha}\\
 $\Delta$ [m/s$^{2}$]& $-0.25$ & disturbance of model\\
 $T$ [s] & $0.02$ & control period \\
\hline\hline
\end{tabular}
\end{center}
\caption{Parameters used in the simulations. \label{tab:params}}
\end{table}

In this section, we use numerical simulations to demonstrate the efficacy of the proposed longitudinal controller in a dynamic traffic environment.
To simulate the real implementation, the proposed controller (\ref{eqn:hdes}-\ref{eqn:ff_cf_term}) is digitally implemented with control period $T=0.02$ [s], while the models (\ref{eqn:long_dynamics},\ref{eqn:car_following_dyn}) evolve continuously over time. In other words, the acceleration command $u$ is updated and then remains constant using zero-order hold in every control loop. Also, Euler integration method is applied to (\ref{eqn:ctrl_nonlinear_rate_lim},\ref{eqn:ctrl_nonlinear_integrator}) at every loop to obtain the acceleration command $u$ and integration error $e$. Other parameters, if not given specifically, will use the values provided in Table~\ref{tab:params}. We remark that these selected control parameters can ensure the stability of the system.

\begin{figure}[!t]
  \centering
  \includegraphics[scale=1.0]{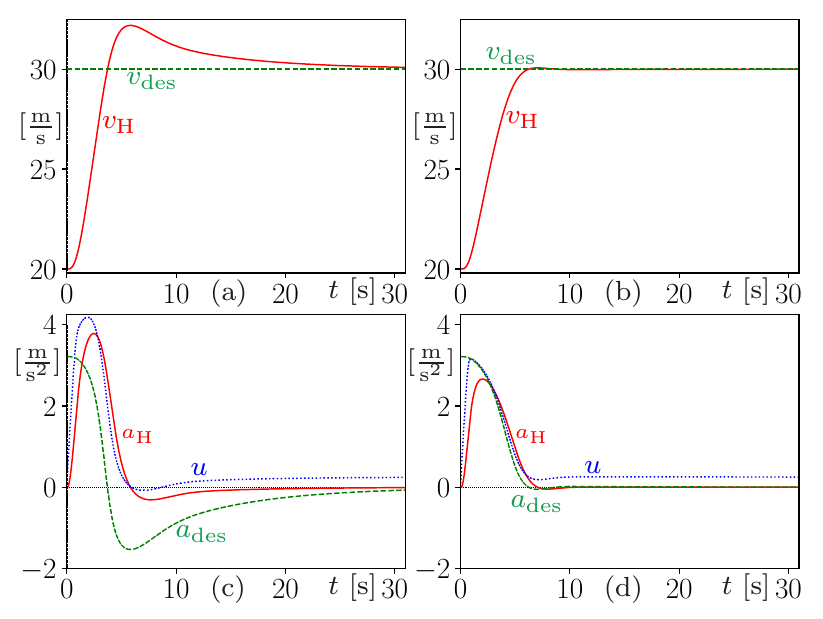}\\
  \caption{Performance comparison on integral controllers in free-driving scenario. (a, c) Velocity and acceleration profiles using normal integral controller \eqref{eqn:ctrl_int_normal}.
(b, d) Velocity and acceleration profiles using nonlinear integral controller \eqref{eqn:ctrl_nonlinear_integrator}. \label{fig:sim_comp_int_wrapper}}
\end{figure}

First we use Fig.~\ref{fig:sim_comp_int_wrapper} to demonstrate the efficacy of the proposed nonlinear integral controller \eqref{eqn:ctrl_nonlinear_integrator} against the normal integral controller \eqref{eqn:ctrl_int_normal} in free-driving scenario. The left column represents the simulation results of the closed-loop system (\ref{eqn:long_dynamics},\ref{eqn:ctrl_nonlinear_rate_lim},\ref{eqn:ctrl_udes},\ref{eqn:ctrl_int_normal},\ref{eqn:vdes_CC},\ref{eqn:ades_CC}) using normal integral controller, while the right column are the results of the system (\ref{eqn:long_dynamics},\ref{eqn:ctrl_nonlinear_PI_rate_lim},\ref{eqn:vdes_CC},\ref{eqn:ades_CC}) using nonlinear integral controller. Panels (a,b) show the time profiles of host vehicle speed $v_{\rm H}$ and desired speed $v_{\rm des}$ using red solid curves and dashed green curves, respectively. In panels (c,d), the red solid curves, the dotted blue curves, and the dashed green curves represent the host vehicle acceleration $a_{\rm H}$, command acceleration $u$, and the desired acceleration $a_{\rm des}$, respectively. The initial values are $v_{\rm H}=20$ [m/s], $v_{\rm des}=30$ [m/s] and $a_{\rm H}=0$ [m/s$^{2}$].

Fig.~\ref{fig:sim_comp_int_wrapper}(a,c) shows that at initial phase, host vehicle accelerates using almost the maximum allowed acceleration ($\sim$ 4 [m/s$^{2}$]) due to integral control with large initial errors. Then the host vehicle continues accelerating until the overshoot of $v_{\rm H}$ over $v_{\rm des}$ is around 3 [m/s]. This is because when $v_{\rm H}$ exceed $v_{\rm des}$, the integration error $e$ is still large and it takes a while for integration error $e$ to decrease such that deceleration efforts are generated. Finally the integration error $e$ converges to the steady-state value (cf.~\eqref{eqn:equil_CC}), and the command acceleration $u$ settles down to balance the disturbance $\Delta$, while the vehicle acceleration $a_{\rm H}$ and vehicle speed $v_{\rm H}$ reaches 0 and $v_{\rm des}$, respectively. Fig.~\ref{fig:sim_comp_int_wrapper}(b,d) indicates that the nonlinear integral controller \eqref{eqn:ctrl_nonlinear_integrator} can significantly improve the performance. Host vehicle acceleration $a_{\rm H}$ only reaches 2.5 [m/s$^{2}$] in the acceleration phase, and then the overshoot of $v_{\rm H}$ over $v_{\rm des}$ only reaches $0.1$ [m/s]. This is because in the initial phase large initial errors are suppressed in the integration \eqref{eqn:ctrl_nonlinear_integrator}, and the nonlinear proportional control \eqref{eqn:ades_CC} dominates the control commands $u_{\rm des}$. Then the integration is gradually fully activated when the difference between $v_{\rm H}$ and $v_{\rm des}$ lies inside $[-\sigma, \sigma]$, and the integration error $e$ converges quickly such that integral part can compensate disturbance $\Delta$. We also observe the following in simulations: i) the overshoot percentage using normal integral controller \eqref{eqn:ctrl_int_normal} increases as the initial errors increase, but that using the proposed nonlinear integral controller \eqref{eqn:ctrl_nonlinear_integrator} remains almost unchanged;
ii) the performance gets worse if the nonlinear controller \eqref{eqn:ades_CC} is replaced with the normal controller \eqref{eqn:ades_CC_lin};
iii) similar results can be obtained for car-following scenarios. These results imply the effectiveness the proposed nonlinear proportional-integral controller in transient response and steady-state response.

\begin{figure*}[!t]
  \centering
  \includegraphics[scale=1.0]{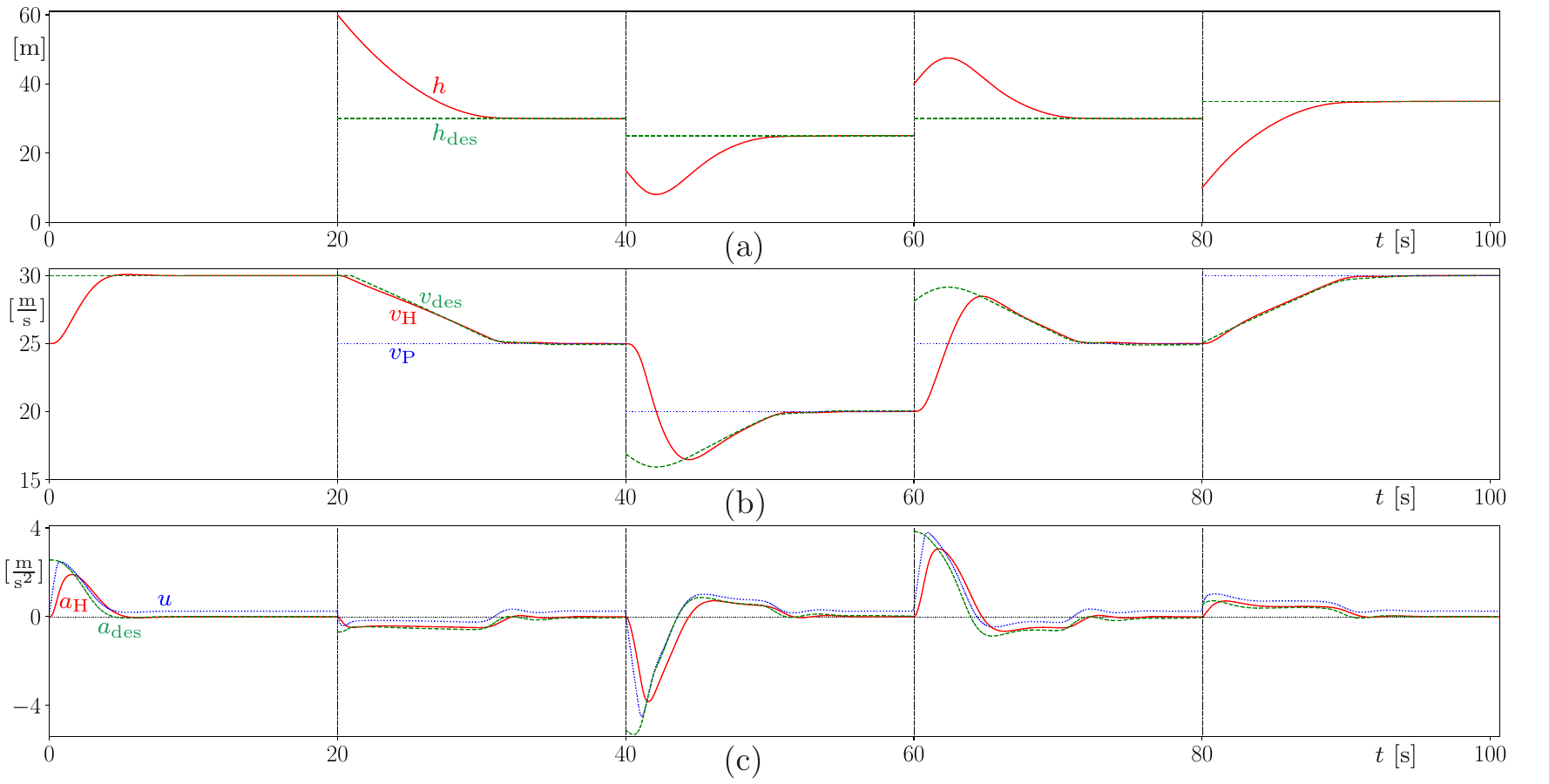}\\
  \caption{Simulation results using controller (\ref{eqn:hdes}-\ref{eqn:ff_cf_term}) when driving scenarios change in a dynamic highway-driving environment. \label{fig:sim_sce1}}
\end{figure*}

Next we simulate a highway-driving situation where preceding vehicles cut in/out, and demonstrate the performance of the proposed longitudinal controller (\ref{eqn:hdes}-\ref{eqn:ff_cf_term}) in such dynamic traffic environment. Fig.~\ref{fig:sim_sce1} shows the simulation results when the driving scenarios changes every $20$ [s] in the following way: i) at $t=0$ [s], host vehicle activates longitudinal controller at $v_{\rm H}=25$ [m/s] with preset maximum speed $v_{\max}=30$ [m/s] in a free-driving scenario; ii) at $t=20$ [s], a slow-moving vehicle ($v_{\rm P}=25$ [m/s]) cuts in far away from the host vehicle ($h=60$ [m]); iii) at $t=40$ [s], another slow-moving vehicle ($v_{\rm P}=20$ [m/s]) cuts in closely ($h=15$ [m]) behind the current preceding vehicle and becomes the new preceding vehicle; iv) at $t=60$ [s], the current preceding vehicle cuts out and the former fast-moving vehicle ($v_{\rm P}=25$ [m/s] and $h=40$ [m]) becomes the preceding vehicle; v) At $t=80$ [s], another fast moving vehicle ($v_{\rm P}=30$ [m/s]) cuts in closely ($h=10$ [m]).

In Fig.~\ref{fig:sim_sce1}(a), the actual inter-vehicle distance $h$ and the desired car-following distance $h_{\rm des}$ are indicated by solid red curves and dashed green curves. Note that in free-driving scenario ($0\le t <20$ [s]), there is no preceding vehicle, resulting in the absence of distance terms. Fig.~\ref{fig:sim_sce1}(b, c) follows the same color scheme as that used in Fig.~\ref{fig:sim_comp_int_wrapper}(a, c) or Fig.~\ref{fig:sim_comp_int_wrapper}(b, d), and utilize dotted blue curves to represent the preceding vehicle speed. It can be seen that in the free-driving scenario, host vehicle accelerates with maximum acceleration around $2$ [m/s$^{2}$], and then settles down around the desired speed $v_{\rm des}=30$ [m/s] with almost no overshoot in about $5$ [s]. As a slow-moving vehicle cuts in far ahead of the host vehicle at $t=20$ [s], the host vehicle decelerates at the comfortable acceleration $a_{\rm com}$ while approaching the preceding vehicle. When the host vehicle speed matches preceding vehicle speed, the inter-vehicle distance reaches desired distance. At $t=40$ [s], when a slow-moving vehicle cuts in closely, the host vehicle is in a safety-critical situation. The controller applies a reasonable deceleration that reaches $-4$ [m/s$^{2}$]. When the host vehicle is slower than the preceding vehicle and collision is mitigated, the controller accelerates the host vehicle with the comfortable acceleration $a_{\rm com}$. Again the distance $h$ reaches desired distance $h_{\rm des}$ once the speed difference decrease to zero. At $t=60$ [s],  the host vehicle starts to accelerate in response to a new far-and-fast-moving preceding vehicle. Once the host vehicle speed is faster than preceding vehicle speed, it coasts down with comfortable acceleration $a_{\rm com}$ while reaching the desired uniform flow state with the preceding vehicle. From $t=80$ [s], the host vehicle starts to accelerate with comfortable acceleration $a_{\rm com}$ as the preceding vehicle changes to another close-but-fast-moving vehicle. In the whole process, we observe that: i) in the presence of disturbance $\Delta$ the desired equilibria are always maintained at steady-state no matter how driving scenarios change; ii) the driving behaviors of the proposed controller are similar to how human drivers react to such scenario changes; and iii) there are no overshoots or oscillations in the transient phase when scenarios change.

\begin{figure*}[!t]
  \centering
  \includegraphics[scale=1.0]{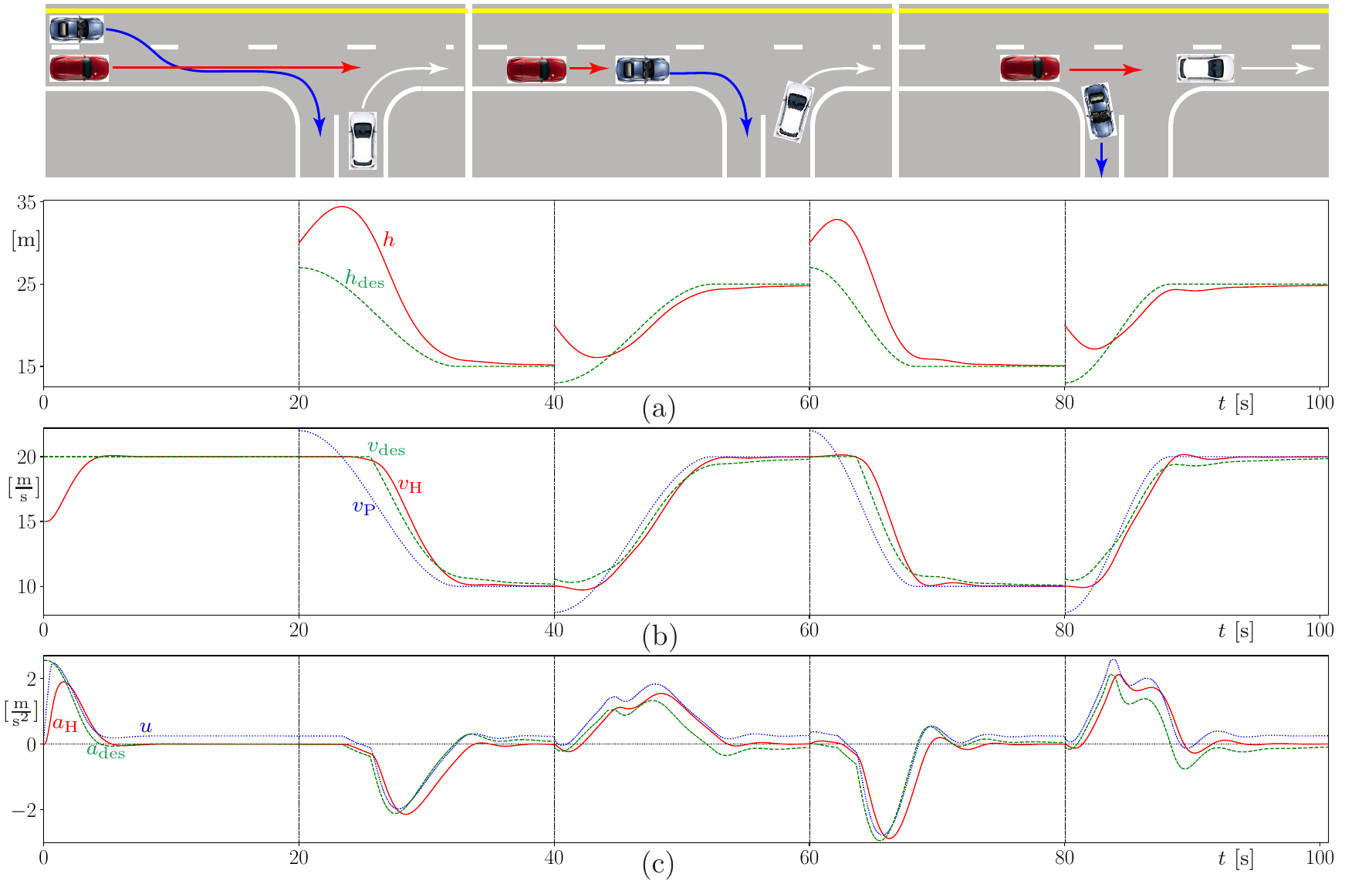}\\
  \caption{Simulation results using controller (\ref{eqn:hdes}-\ref{eqn:ff_cf_term}) when driving scenarios change in a dynamic local-driving environment. \label{fig:sim_sce2}}
\end{figure*}

We also run simulations with the proposed longitudinal controller (\ref{eqn:hdes}-\ref{eqn:ff_cf_term}) in a local-driving situation where preceding vehicles cut in/out and accelerate/decelerate. Simulation results are shown in Fig.~\ref{fig:sim_sce2}, where the same color scheme is used as that in Fig.~\ref{fig:sim_sce1}. The first row depicts the following situation: i) at $t=0$ [s], host vehicle activates longitudinal controller at $v_{\rm H}=15$ [m/s] with preset maximum speed $v_{\max}=20$ [m/s] in a free-driving scenario; ii) at $t=20$ [s], a fast-moving vehicle ($v_{\rm P}=22$ [m/s]) cuts in at $h=30$ [m], and starts to decelerate and prepare for the right turn to exit; iii) at $t=40$ [s], the former preceding vehicle make a right turn to the branch road and the preceding vehicle changes to a new slow-moving-and-accelerating vehicle ($v_{\rm P}=8$ [m/s]) at $h=20$ [m] that has just merged from the branch; iv) at $t=60$ [s], the scenario change is the same to that happens at $t=20$ [s] except that the preceding vehicle decelerates faster; and v) at $t=80$ [s], the scenario change is the same to that happens at $t=40$ [s] except that the preceding vehicle accelerates faster. The results indicate that the proposed controller allows the host vehicle to reasonably respond to changes in preceding vehicle states and settles down around the desired equilibrium in the presence of disturbance when driving scenarios change.

\begin{figure*}[!t]
  \centering
  \includegraphics[scale=1.0]{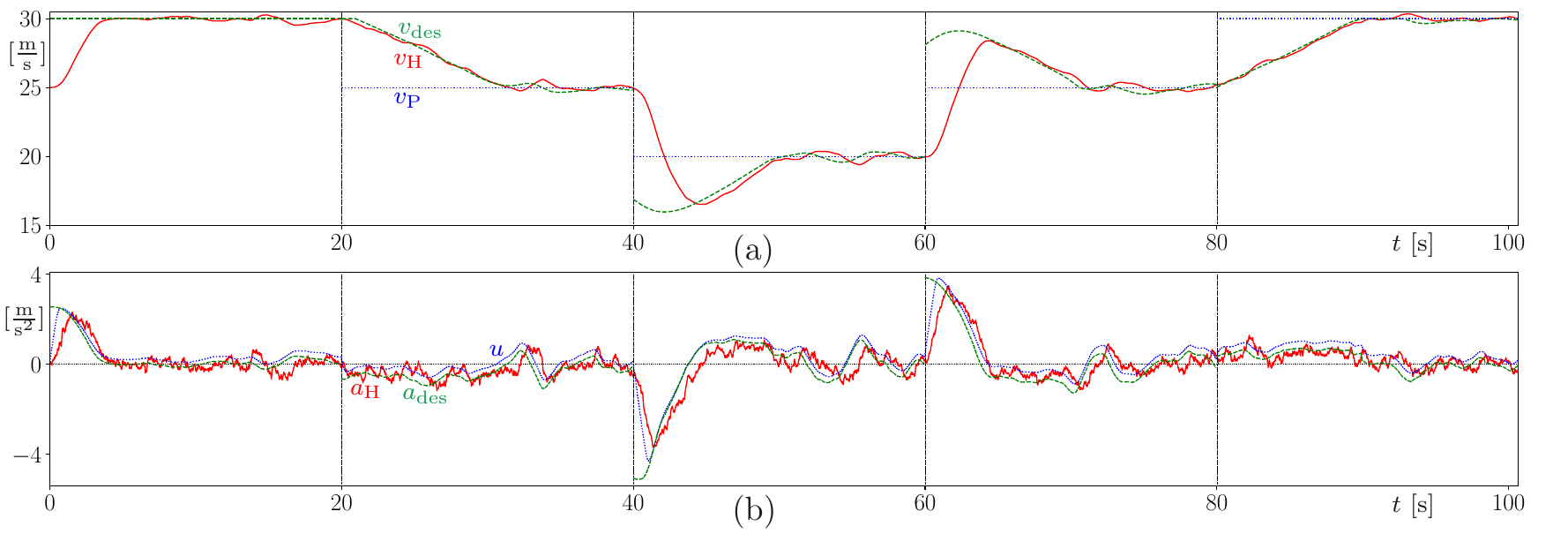}\\
  \caption{Simulation results in the same highway-driving environment as Fig.~\ref{fig:sim_sce1} when the disturbance $\Delta$ varies randomly  according to Gaussian distribution $\mathcal{N}(-0.25, 0.25^{2})$. \label{fig:sim_sce3}}
\end{figure*}

Fig.~\ref{fig:sim_sce1} and \ref{fig:sim_sce2} demonstrate the efficacy of the proposed controller in handling both transient response and steady-state response in scenario changes when the disturbance is constant $\Delta=0.25$ [m/s$^{2}$]. Indeed, in a dynamic traffic environment, the disturbance is also changing dynamically due to variations in road roughness, road grade, air density, etc. Fig.~\ref{fig:sim_sce3} simulates the same driving situation as Fig.~\ref{fig:sim_sce1} when the disturbance varies randomly and satisfies the Gaussian distribution $\mathcal{N}(-0.25, 0.25^{2})$. We only show the speed and acceleration profiles in the remainder of this section since the differences in the distance profile against Fig.~\ref{fig:sim_sce1}(a) are so tiny to be noticeable. Similar behaviors are observed in Fig.~\ref{fig:sim_sce3} as those in Fig.~\ref{fig:sim_sce1}. Regardless of the dynamic disturbance, the proposed controller properly handles transient response when scenarios change, and gradually reaches the desired equilibrium in new driving scenarios without ``overreaction". The variations of host vehicle speed $v_{\rm H}$ around the desired values $v_{\rm des}$ implies that the proposed controller are capable of maintaining desired equilibrium in the presence of dynamical disturbance.

\begin{figure*}[!t]
  \centering
  \includegraphics[scale=1.0]{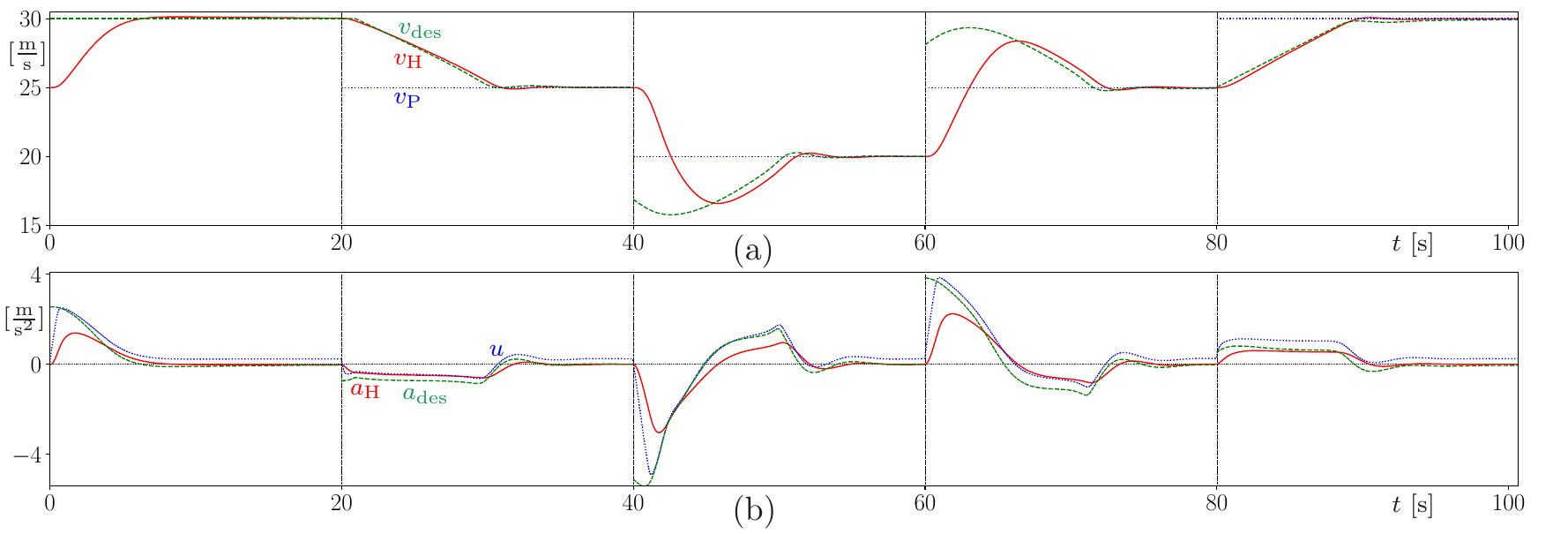}\\
  \caption{Simulation results in the same highway-driving environment as Fig.~\ref{fig:sim_sce1} when there exists a $30\%$ under-estimation error on vehicle mass. \label{fig:sim_sce4}}
\end{figure*}

\begin{figure*}[!t]
  \centering
  \includegraphics[scale=1.0]{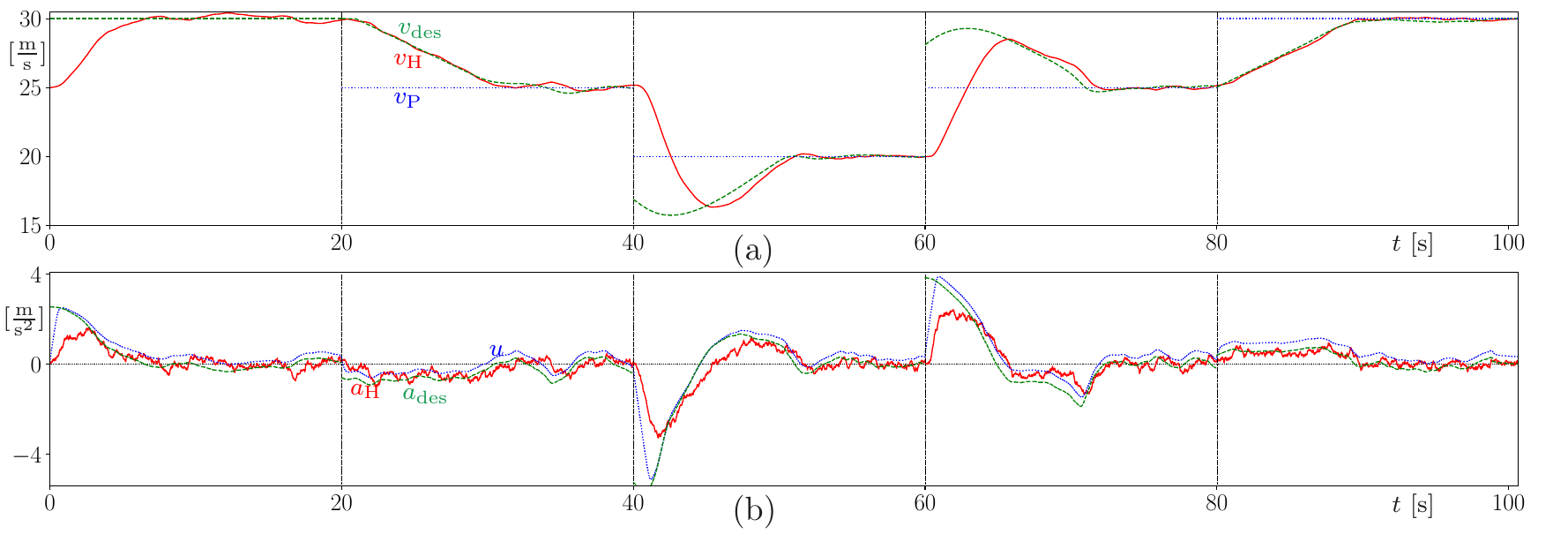}\\
  \caption{Simulation results in the same highway-driving environment as Fig.~\ref{fig:sim_sce1} when $\alpha_{1}=0.7$ and the disturbance $\Delta$ varies randomly according to Gaussian distribution $\mathcal{N}(-0.25, 0.25^{2})$. \label{fig:sim_sce5}}
\end{figure*}

Fig.~\ref{fig:sim_sce4} simulates the same driving environment as Fig.~\ref{fig:sim_sce1} when the estimation on vehicle mass is not accurate. This is a common problem because low-level controllers (such as \eqref{eqn:low_level_trq_ctrl}) convert command acceleration $u$ to desired torque $T_{\rm des}$ based on nominal vehicle mass $\hat{m}$. However, vehicle mass varies when the number of passengers or the weight of cargo change.  Fig.~\ref{fig:sim_sce4} shows the results when there exists a $30\%$ under-estimation error on vehicle mass, i.e., $\alpha_{1}=0.7$. We observe the following: i) similar to the behaviors in Fig.~\ref{fig:sim_sce1}(b,c), the controller handles transient response properly in scenario changes and maintains desired equilibrium state at steady-state regardless of the estimation error of vehicle mass; ii) due to under-estimate of vehicle mass, host vehicle acceleration $a_{\rm H}$ is slightly smaller than that in Fig.~\ref{fig:sim_sce1}(a); and iii) there is a slightly larger difference between $a_{\rm H}$ and $a_{\rm des}$ in the transient phase due to inadequate torque $T$ using under-estimated mass. We remark that similar behaviors can be observed if vehicle mass is over-estimated, and host vehicle acceleration $a_{\rm H}$ will be slightly larger.

Fig.~\ref{fig:sim_sce5} shows the simulation results when dynamic traffic environment, dynamic disturbance and estimation error on vehicle mass are all considered. The driving environment is the same as Fig.~\ref{fig:sim_sce1}, the dynamic disturbance is the same Gaussian-distributed random signal used in Fig.~\ref{fig:sim_sce3}, and the estimation error on vehicle mass is still $30\%$ less as that in Fig.~\ref{fig:sim_sce4}. We observe the same nice performance of the proposed controller in the presence of all these effects. The proposed controller ensures reasonable behaviors in the transient response while approaching desired equilibria when scenarios change, and then maintains desired steady-state equilibrium afterwards.

In this section, we used simulations to demonstrate the efficacy of the proposed controller in dynamic traffic environments in the presence of dynamic disturbances and estimation errors on vehicle mass. It is shown that the proposed controller is capable of handling transient response reasonably as human drivers do while reaching the new desired equilibrium state when driving scenarios change, regardless of the effects of disturbances and estimation errors. It is a common architecture in automated vehicles that longitudinal controllers ensure stability of steady-state equilibrium, while planning algorithms generate trajectories for those controllers to guarantee transient response. Simulation results indicate that the proposed controller can work properly even without the usage of planning algorithms, which can solve the issues caused by coupling nature between planning and control. This is because the proposed controller implicitly includes the concept of planning in the desired speed $v_{\rm des}$ inspired by human-driving behaviors, which depends on preceding vehicle speed $v_{\rm P}$, inter-vehicle distance $h$ and ultimate desired distance $h_{\rm des}$. The control algorithm guarantees that the vehicle follows desired speed $v_{\rm des}$ in the transient response while approaching steady-state equilibria. In comparison, in the normal design planning algorithms generate trajectories of $v_{\rm des}$ and $h_{\rm des}$ as explicit functions of time $t$ based on nominal vehicle models, preceding vehicle speed $v_{\rm P}$ and inter-vehicle distance $h$. The proposed controller combines the planning and control techniques together by ensuring not only the stability of the desired equilibria, but also that of the desired ``trajectory". Also, by utilizing the nonlinear constraint on command changing rate and nonlinear proportional-integral control technique, reasonable behaviors are guaranteed while approaching the desired ``trajectory" at the time instant when driving scenarios change.

\section{Conclusion\label{sec:conclusion}}
In this paper we proposed a comprehensive longitudinal controller that integrates controllers for free-driving scenarios and car-following scenarios. This controller can ensure reasonable and smooth transient response in scenario changes, and also guarantee stabilizability of desired equilibria in different driving scenarios in the presence of dynamical disturbances. Simulations are performed under different driving scenario changes, and the results indicate responsive and reasonable actions of controlled vehicle to adapt to new scenarios without usage of planning algorithms. This implies that tasks of planning algorithms can be significantly reduced if controllers can properly handle transient response and steady-state response in scenario changes. Also, the coordination issue in the integration of planning and control can be resolved. Future research directions may include field experiments, time delays inherent in sensing, optimal design, etc.


\bibliographystyle{IEEEtran}

\begin{IEEEbiography}
[{\includegraphics[width=1in,height=1.25in,clip,keepaspectratio]{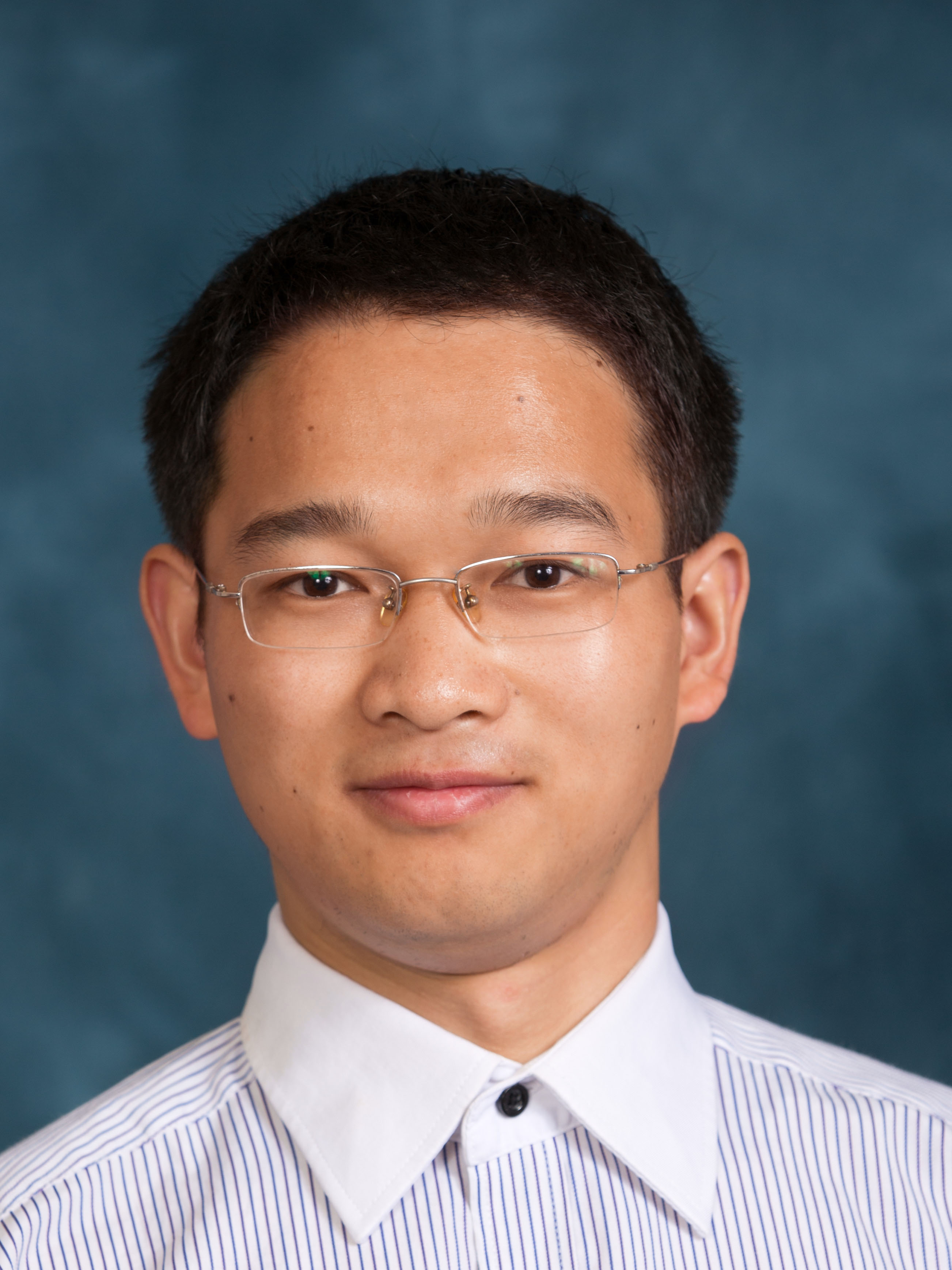}}]{Wubing B. Qin}
received his BEng degree in School of Mechanical Science and Engineering from Huazhong University of Science and Technology, China in 2011, and his MSc degree and PhD degree in Mechanical Engineering from the University of Michigan, Ann Arbor in 2016 and 2018, respectively. Curretly he is an independent researcher without affiliations. His research focuses on dynamics, control theory, connected/automated vehicles, ground robotics, mechatronics and nonlinear systems.
\end{IEEEbiography}

\end{document}